\documentclass[twocolumn,showpacs,prl,aps,floats,amssymb]{revtex4}
\usepackage{graphicx}

\begin{document}

\title{Breakdown of Fermi-liquid theory in a cuprate superconductor}

\author{R.W. Hill,
Cyril Proust$^\dagger$ and Louis Taillefer}

\address{Canadian Institute for Advanced Research}

\address{
Department of Physics, University of Toronto, Toronto,
Ontario M5S 1A7, Canada}

\author{P. Fournier$^\ddagger$ and R.L. Greene}

\address{Center for Superconductivity Research\\ Department of Physics, 
University of Maryland, College Park, Maryland 20742, USA}

\date{December 13, 2001}

\begin{abstract}

The behaviour of electrons in solids is remarkably well described by Landau's Fermi-liquid
theory, which says that even though electrons in a metal interact they can still be treated
as well-defined fermions, called ``quasiparticles''.
At low temperature, the ability of quasiparticles
to transport heat is strictly given by their
ability to transport charge, via a universal relation known as the Wiedemann-Franz law,
which no material in nature has been known to violate.
High-temperature superconductors have long been thought to fall outside the realm of 
Fermi-liquid theory, as suggested by several anomalous properties, but this has yet to
be shown conclusively. 
Here we report on the first experimental test of the Wiedemann-Franz law in a 
cuprate superconductor, (Pr,Ce)$_2$CuO$_4$. 
Our study reveals a clear departure from the universal law and provides 
compelling evidence for the breakdown of Fermi-liquid theory in high-temperature 
superconductors.

\end{abstract}


\maketitle

\noindent
{\bf INTRODUCTION}\\

Landau's Fermi-liquid theory is the definitive theory of electrons in metals \cite{Landau}, or
more generally fermions in condensed matter, and a major landmark of 20th century physics.
For example, it is the necessary foundation for the theory of superconductivity by Bardeen, Cooper and Schrieffer
(BCS) \cite{BCS}.
In essence, it says that
 even in the presence of
interactions the low-energy excitations of a system of mobile
fermions can still be described in terms of
well-defined
fermionic particles, called ``quasiparticles'', with charge $e$, spin $\frac{1}{2}$ and
a mass $m^{\star}$, the latter being renormalized by interactions.
The Wiedemann-Franz (WF) law is one of the basic properties of a Fermi liquid, reflecting
the fact that the ability of a quasiparticle to transport heat is the same
as its ability to 
transport charge, provided it cannot lose energy through collisions.
Reported as an empirical observation by Wiedemann and Franz in 1853
\cite{WFL}, the law states that 
the heat conductivity $\kappa$ and
the electrical conductivity $\sigma$ of a  
metal are related by a universal constant:
\begin{equation}
\frac{\kappa}{\sigma T} = \frac{\pi^2}{3} ~ (\frac{k_B}{e})^2 \equiv L_0
\end{equation}
where $T$ is the absolute temperature,
$k_B$ is Boltzmann's constant and $L_0=
2.45 \times 10^{-8}$~W~$\Omega$~K$^{-2}$ 
is Sommerfeld's value for the 
Lorenz ratio
$L \equiv \kappa/\sigma T$ \cite{Sommerfeld}.

The linear power of temperature in Eq.~(1) comes from the linear temperature dependence
of the fermionic specific heat, through the 
relation between heat transport and heat capacity. In kinetic theory,
\begin{equation}
\kappa = \frac{1}{3} ~ c ~ v~ l
\end{equation}
where $c$ is the specific heat of the carriers, $v$ is their average velocity and $l$ their mean free
path.
Provided the mean free path between collisions
is independent of temperature, as it is for the strictly elastic processes that operate at $T \to 0$,
the heat conduction has the same $T$ dependence as the specific heat. So for electrons,
it is also linear in $T$, and given by Eq.~(1).
By contrast, boson-like excitations such as phonons or magnons give rise to a
low-temperature specific heat and a heat conductivity as $T \to 0$ which
are proportional to $T^3$.

Theoretically, electrons are predicted to obey the WF law at $T \to 0$ 
in a very wide range of environments \cite{Kearney}:
in both three dimensions or two dimensions (but not strictly
in one dimension), for any strength of disorder and interaction
(in the scaling theory of disorder and interaction) \cite{Castellani},
scattering \cite{Chester} and magnetic field \cite{Smrcka}.
Experimentally, the WF law does appear to be universal at $T \to 0$: 
no material has been reported
to violate 
it. (Note that a departure from Eq.~(1)
has in fact been reported for silver \cite{Pobell}, but is in
contradiction with other measurements \cite{Cu1,Cu2}, 
including our own
\cite{Ag-test}.)
This is true not only of simple metals like copper \cite{Cu1,Cu2}, 
but also of systems with strong
electron correlations -- such as the heavy fermion compounds CeAl$_3$ \cite{Ott},
CeCu$_6$ \cite{Amato} and UPt$_3$ \cite{Suderow} --
or highly anisotropic conduction --
such as the quasi-2D system
Sr$_2$RuO$_4$ \cite{Tanatar} (a ruthenate which is isostructural to the cuprate La$_2$CuO$_4$), 
the quasi-1D organic conductor (TMTSF)$_2$ClO$_4$ \cite{Belin}, or 
the 2D electron gas in a MOSFET \cite{Pepper}. 
Even in the presence of so-called
non-Fermi liquid behaviour, where the electrical
resistivity is seen to deviate from the standard $T^2$ dependence,
the WF law still holds at $T \to 0$, 
as observed in CeNi$_2$Ge$_2$ \cite{Kambe}. 
All this strongly suggests that the ground state
of every metal investigated thus far is a Fermi liquid. Of course, if the metallic
state gives way to a superconducting state
at low temperature, then the WF law is entirely violated, as the charge is no longer 
transported by fermions but by Cooper pairs, which carry no entropy.

In 1986, high-temperature superconductivity was discovered in a class of oxides called cuprates,
and 
fifteen years hence the fundamental question about these materials still remains:
are they host to a new state of matter?
In particular, are they Fermi liquids in their ground state, once superconductivity is removed
(either by application of a magnetic field or by doping)?
Anderson proposed early on
\cite{Anderson} that the cuprates are indeed fundamentally different from other
metals in that they are Mott insulators.
When a small number of electrons or holes are doped into the CuO$_2$ planes of their crystal
structure, these materials
can conduct charge reasonably well and also form a superconducting state.
It is believed that the 
basic excitations of a doped Mott insulator are not fermionic quasiparticles 
as in other metals -
with charge, spin and heat all carried by one and the same particle.
Instead, the electron is thought to ``fractionalize'' into a neutral
spin-$\frac{1}{2}$ 
fermion called a ``spinon'' and a spinless charge-$e$ 
boson called a ``holon'', or a ``chargon'' \cite{Kivelson}, a phenomenon
called ``spin-charge'' separation.

Even though this bold proposal for a fundamentally new state of matter continues to be at the heart 
of several current theories of the cuprates \cite{Fisher}, 
there has been little experimental evidence for
such spin-charge separation. On the contrary, for example, 
a BCS/Fermi-liquid theory of $d$-wave quasiparticles
appears to work rather well in the superconducting state of 
hole-doped cuprates, at least at low energy and near 
optimal doping
\cite{Millis}.
Conceptually, a simple way to investigate the possibility of spin-charge
separation
is to measure the properties
of charge transport and compare them to those of either spin or entropy
transport.
In this Article, we report on the first test of this kind, where both the 
charge conductivity $\sigma$ and the heat conductivity $\kappa$ of a cuprate
superconductor are measured in the normal state
at $T \to 0$. {\sl We observe a good, metallic-like nearly temperature-independent
charge transport at low temperature without the corresponding 
heat transport expected from the WF law}.
The normalized
Lorenz number $L/L_0$ goes from being much
less than unity at $T \to 0$ to being greater than unity above 0.2 K. 
This is highly suggestive of spin-charge separation
and a compelling demonstration that the ground state of cuprates  
can fall outside the realm of Fermi-liquid theory.\\
\\
{\bf THE MATERIAL}\\

The cuprate material used for this study is Pr$_{2-x}$Ce$_x$CuO$_{4-y}$ (PCCO), with a doping concentration
of $x=0.15$, which is roughly 
the optimum value for maximizing the critical temperature for superconductivity, where
$T_c \simeq$~20 K.
The addition of $x$ Ce atoms on Pr sites adds $x$ electrons to the CuO$_2$ planes of the parent
insulating compound, the Mott insulator Pr$_2$CuO$_4$. 
PCCO is the electron-doped analog of the hole-doped
material La$_{2-x}$Sr$_x$CuO$_{4-y}$ (LSCO), wherein the addition of $x$ Sr atoms on La sites
removes $x$ electrons from (adds $x$ holes to)
the CuO$_2$ planes.
In LSCO, the optimum doping is also at approximately $x=0.15$, with a maximum $T_c \simeq$~40 K.
For reasons that are not clear, both materials have critical temperatures that are significantly
lower than those of several other cuprates, such as YBa$_2$Cu$_3$O$_{7-\delta}$ (Y-123), 
Bi$_2$Sr$_2$CaCu$_2$O$_{8+\delta}$ (Bi-2212)
and Tl$_2$Ba$_2$CuO$_{6+\delta}$ (Tl-2201), which all have a maximal $T_c$ around 90 K.
Nevertheless, all cuprates share 
the same basic crystal structure, namely that of a stack of CuO$_2$ planes, and the
same generic dependence on doping.
The technical advantage of LSCO and PCCO is that
the magnetic field needed to destroy superconductivity is correspondingly lower, with upper critical
field values of $\simeq$~50 T and $\simeq$~10 T, respectively. Because these fields can be achieved in the 
laboratory, it is possible to investigate their low-temperature transport properties in the normal  
state, such as was done for charge conductivity
by Boebinger {\it et al.}
on LSCO \cite{Boebinger} and by Fournier {\it et al.} on PCCO \cite{Fournier}.\\
\\
{\bf CRYSTALS}\\

The single crystal of PCCO used in this study was grown with a flux technique described elsewhere
\cite{Growth1,Growth2}, using a Al$_2$O$_3$
crucible and an oxygen reduction treatment at 1000 $^{\circ}$C.
The concentration of cerium was fixed at $x=0.15$.
Contacts were made with silver epoxy, diffused at 500 $^{\circ}$C for 1 hour, and were used to measure
both electrical resistivity and thermal conductivity.
The electrical resistance of contacts was typically 
$\simeq 1~\Omega$.
The crystal was in the shape of a platelet, 37 $\mu$m thick and 720 $\mu$m wide, with 1.0 mm 
separation between voltage contacts. 
The superconducting transition temperature $T_c$ was obtained from three different measurements:
as the end of the drop in resistivity, the end of the diamagnetic drop in susceptibility 
and the onset 
of a small peak in thermal conductivity. 
In all cases, $T_c = 20$~K to within less than $\pm 0.5$~K.
The width of the transition is 6~K in resistivity, 3~K in magnetisation and about 1~K in 
thermal conductivity. The broad resistive transition, seen both as a function of temperature and 
magnetic field, is most likely due to an inhomogeneous distribution of oxygen near the surface.
The narrower transition
obtained in thermal conductivity suggests that this inhomogeneity is confined to a negligible volume 
of the sample, and therefore does not affect bulk measurements such as heat transport.
Note that all results presented in this paper were reproduced on 
several other single crystals grown in similar conditions, with comparable $T_c$ and resistivity.\\

\begin{figure}[b]
\centering
\resizebox{\columnwidth}{!}{\includegraphics*{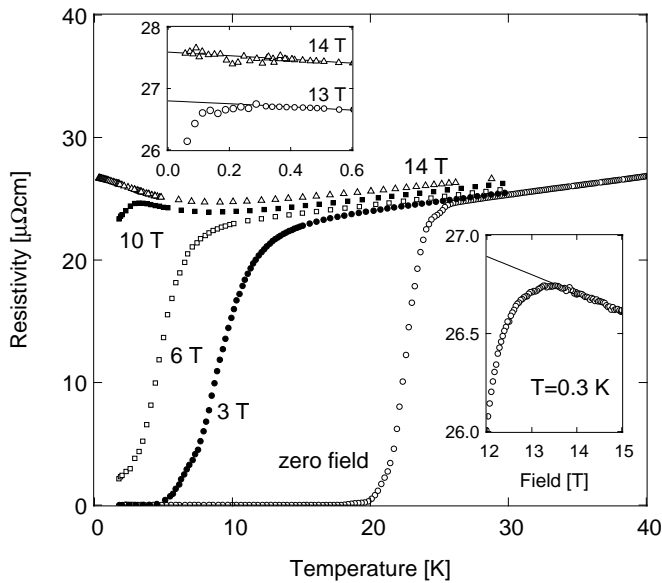}}
\vspace{4pt}
\caption{Electrical resistivity of PCCO vs temperature
for a current in the basal plane at different values of magnetic field applied
normal to the plane. {\it Upper inset}: zoom of low temperature dependence at 13 T and
14 T, the latter being shifted upwards by 0.8~$\mu \Omega$~cm for clarity. Lines 
are fits to the data below 1.0~K. {\it Lower inset}: 
Resistivity vs magnetic field at $T=0.3$~K, showing that 
the crystal has fully completed its transition to the normal state by 13.5~T. 
The line is a linear fit 
to the negative magnetoresistance beetween 13.5 and 15~T.}
\label{Fig2}
\end{figure}

\noindent
{\bf CHARGE TRANSPORT}\\

The temperature dependence of the electrical resistivity is shown in Fig.~\ref{Fig2} 
for different magnetic fields applied
parallel to the $c$-axis of the tetragonal crystal structure.
The current is made to flow in the CuO$_2$ planes, 
where the conduction is some
10$^4$ times better than 
perpendicular to the planes (at room temperature).
Above 25~K, the temperature dependence is not linear as in most optimally-doped cuprates,
 but follows a power law $a +
b T^{\alpha}$ with 
$\alpha \simeq 1.7$, 
and has a magnitude characteristic of a reasonably good metal: $a \simeq  25~\mu \Omega$~cm 
and $\rho$(300~K) $\simeq$ 140 ~$\mu \Omega$~cm. 
The effect of a magnetic field is to rapidly
suppress the superconducting transition initially, but then more slowly at high fields, giving 
rise to a curve $T_c(H)$ that has strong positive curvature. 
From the 
magnetoresistance at 0.3 K shown in the lower inset, one can see 
that the entire sample has completed its
transition to the normal state by 13.5 T.
Note, however, that the bulk of the sample is no longer superconducting above 8 T, the field
beyond which the low-temperature
heat transport ceases to evolve (see Fig.~\ref{Fig3} below).

The overall behaviour of the resistivity in PCCO 
is similar to that of hole-doped cuprates when overdoped.
Indeed, when doped to have a $T_c$ of 15~K, Tl-2201 has a resistivity which
goes as $T^{1.75}$ with values of 
10~$\mu \Omega$~cm at low temperature 
and 180~$\mu \Omega$~cm at room temperature, and a resistive critical field with a strong upward curvature
reaching a value of 16 T at $T \to 0$ 
\cite{Mackenzie}.
One difference is the slight upturn seen at low temperature in PCCO and not in Tl-2201.
The resistivity in the normal state (at 14 T) increases by about 10\% in going from
8~K down to 
0.3~K (see Fig.~\ref{Fig2}).  
Similar but much larger upturns were observed
in Nd$_{2-x}$Ce$_x$CuO$_4$
(NCCO) \cite{Harus,Fournier2} and PCCO \cite{Fournier2} at lower carrier concentrations,
and attributed to weak localization.
In support of this interpretation is the suppression of the upturn by a transverse
magnetic field, also observed in our samples (see inset of Fig.~\ref{Fig2}).
This
negative magnetoresistance 
eventually flattens to a constant beyond 25 T, equal to 25.2~$\mu \Omega$~cm at 0.5 K \cite{Hill}.

In summary, once the superconducting order has been completely suppressed throughout the sample, 
namely in a magnetic field larger
than 13.5 T, the charge transport of this cuprate material at low temperature
is that of a fairly 
good quasi-two-dimensional
metal, with a residual resistivity $\rho(T \to 0) = \rho_0 = 26.8~ 
\mu \Omega$~cm in 14 T, 
which corresponds
to a conductivity
per CuO$_2$ plane of 60~$e^2/h$ ({\it i.e.} $k_F l \simeq 60$).\\
\\
\noindent
{\bf HEAT TRANSPORT}\\

Fifteen years after the discovery of high-temperature superconductivity,
very little is known about the normal state
of cuprates at low temperature.
The ability to suppress superconductivity in optimally-doped PCCO
with a field of only 14 T
provides a rare opportunity to shed light on the fundamental nature of the electron system 
in these
materials. In the present study, this 
was done by measuring the transport of heat
as $T \to 0$. The thermal conductivity $\kappa(T)$ of PCCO is shown in Fig.~\ref{Fig3} 
at temperatures below
0.3~K, for heat flowing parallel to the CuO$_2$ planes. 
The data is plotted as $\kappa/T$ vs $T^2$ in order to easily separate the contributions of phonons 
and electrons, using the fact that
phonons are expected
to have a heat
conductivity 
which varies as $T^3$ when their mean
 free path reaches the size of the crystal at sufficiently
low temperature. A straight line extrapolating to zero in $\kappa/T$ vs $T^2$ is
indeed the asymptotic ($T \to 0$) behaviour
 observed in insulators \cite{Thacher}, including the Mott insulator
YBa$_2$Cu$_3$O$_6$ ({\it i.e.} Y-123 without
any doped carriers) \cite{Taillefer}. This is also the behaviour of conventional ($s$-wave) 
superconductors.
The absence of a residual
linear term  $\kappa_0/T$, defined as the value of $\kappa/T$ as $T \to 0$,
  means that there are no mobile fermionic
excitations at low temperature. In insulators, this is due
to the lack of electronic carriers, while in superconductors it is because electrons
have all formed Cooper pairs at $T << T_c$. The heat 
is then conducted entirely by phonons.
In contrast, a residual linear term 
is observed
when fermionic excitations are present, as in the following three
instances:
1) in metals, with a residual linear term given precisely by the WF law;
2) in the vortex state of
superconductors, where a magnetic field induces a residual linear term
by generating 
delocalized quasiparticles
\cite{Boaknin2};
3) in $d$-wave superconductors, where zero-energy quasiparticles are induced by impurity scattering
\cite{Durst}.\\
\begin{figure}[t]
\centering
\resizebox{\columnwidth}{!}{\includegraphics*{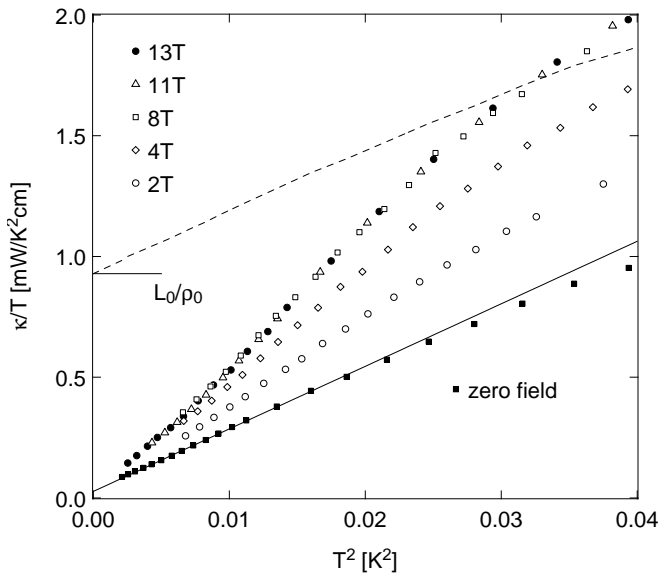}}
\vspace{4pt}
\caption{Thermal conductivity of PCCO for a heat current in the basal
plane, plotted as $\kappa/T$ vs $T^2$,
at different values of the magnetic
field applied normal to the plane. 
The solid line is a linear fit to the zero-field data below
130 mK.
The dashed line shows the behaviour of a Fermi-liquid with the residual resistivity
$\rho_0$ of this sample, calculated as the sum of a constant electronic
linear term via Eq.~(1) and a phonon conductivity
given by the zero-field data. 
}
\label{Fig3}
\end{figure}

\noindent
{\bf Superconducting state}\\

In the absence of a magnetic field,  
there is only a very small residual linear term in the thermal conductivity of PCCO.
From data on four different crystals, we find $\kappa_0/T = 0.03 - 0.07$~ mW K$^{-2}$ cm$^{-1}$. 
This is considerably smaller than the
residual linear term observed in three hole-doped cuprates, at 
optimal doping: 
$\kappa_0/T$ = 0.14, 0.15 and 0.11 mW K$^{-2}$ cm$^{-1}$, in
Y-123 \cite{Taillefer,Chiao}, Bi-2212 \cite{Chiao2,Aubin} and LSCO \cite{Taillefer2},
respectively.
Within BCS theory applied to a $d$-wave superconductor, this fermionic residual heat conduction
is expected, arising from zero-energy quasiparticles 
induced by impurity scattering near the nodes in the $d_{x^2-y^2}$ gap function \cite{Durst}.
In the case of Bi-2212, the 
excellent quantitative agreement between theory and experiment has been viewed as a strong 
validation of a Fermi-liquid quasiparticle picture for the superconducting state
\cite{Millis,Chiao2}, 
at least at optimal doping and low energies.

Several tentative explanations
may be offered for the weakness of the residual
linear term in PCCO.
The first possibility
is the localization of $d$-wave quasiparticles,  
investigated by several authors \cite{Lee}.
It was invoked recently as a possible cause for the 
absence of a residual linear term in YBa$_2$Cu$_4$O$_8$
(Y-124) \cite{Hussey}. However, it
is not clear why localization should occur in Y-124 and not in Y-123, 
given that these two cuprates have comparable
structure and properties. 
It seems more
plausible in PCCO, because of the highly two-dimensional nature of its conduction --
some 100 times more anisotropic than either Y-124 or LSCO.
On the other hand, Bi-2212 is equally anisotropic, yet shows delocalized quasiparticles.
A second scenario is the absence of nodes in the gap structure of PCCO.  
This 
would eliminate 
any 
residual linear term (at least in zero magnetic field). The symmetry of the order parameter
need not be $s$-wave, but could 
have the form $d+ix$, where $x$ is a subdominant component of either $s$ or $d_{xy}$ symmetry.
The possibility of a $T=0$ transition in the order parameter from $d$ to $d+ix$
as a function of doping has been
raised in the context of a 
possible quantum critical point in the phase diagram of cuprates \cite{Sachdev}.
For PCCO, however, 
this would seem to be in conflict with evidence for a rather pure $d_{x^2-y^2}$ symmetry as 
determined in a recent tricrystal experiment \cite{Tsuei}. The other difficulty
is that, as we shall see below, 
a magnetic field normal to the CuO$_2$ planes never induces a residual
linear term, implying that the minimum gap would have to be a sizable fraction of the maximum
gap ({\it i.e.} $x$ comparable to $d$). This would have to be reconciled with the penetration
depth data \cite{lambda1,lambda2}.
A third possibility, proposed recently by Granath {\it et al}. \cite{Granath}, is the removal
of the nodes as a result of static stripe ordering.\\
\\
\noindent
{\bf Normal state}\\

As seen in Fig.~\ref{Fig3}, 
the basic effect of a magnetic field is to rapidly increase the thermal conductivity 
of PCCO (at finite temperature). 
Given that the phonon conductivity cannot exceed
its value at zero field (where the phonon mean free path has reached its maximum value), this increase
must be due
to excitations of electronic origin. 
The field-induced conduction (over and above the zero-field curve)
grows rapidly (and approximately linearly) with field at low fields 
to eventually saturate at
high fields, where $\kappa$ is absolutely constant above 8 T.
This is compelling evidence that the thermodynamic upper
critical field in this sample is at most
8 T and the bulk of the crystal is in the normal state beyond that. 
Indeed, in all known superconductors, whether $s$-wave or $d$-wave, the electronic
thermal conductivity is always affected by the magnetic field when in the vortex state,
and seen to saturate above $H_{c2}$ (see for example Ref.~\cite{Belin2}).
\begin{figure}[t]
\centering
\resizebox{\columnwidth}{!}{\includegraphics*{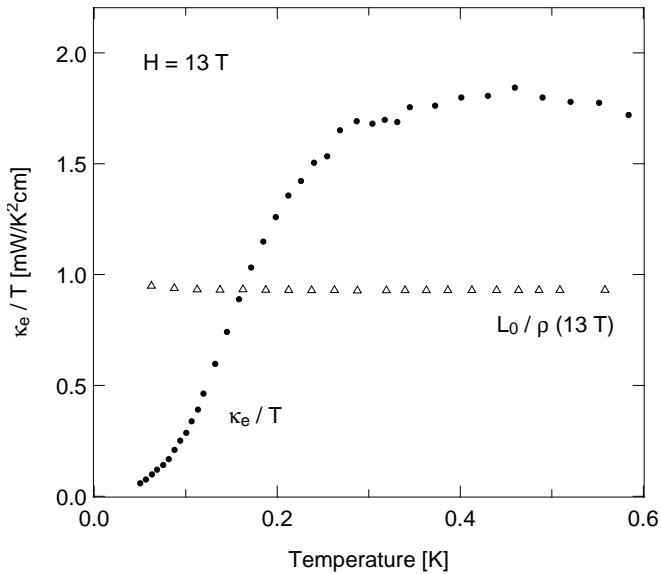}}
\vspace{4pt}
\caption{Comparison of charge conductivity $\sigma(T)=1/\rho(T)$, plotted
as $L_0/\rho(T)$ (open triangles), and electronic 
heat conductivity $\kappa_e$, plotted as
$\kappa_e/T$ (full circles), as a function of temperature in the normal state at $H=13$~T.
The electronic contribution to the heat conduction
is the difference between the measured $\kappa$(13 T) and the phonon contribution
$\kappa_{ph}$(13 T), estimated in the text.
In a Fermi liquid, the curve of 
$\kappa_e/T$ would lie precisely on top of the data for
$L_0/\rho$. Below 0.15~K, $\kappa_e \sim T^{3.6}$.
}
\label{Fig4}
\end{figure}
\noindent
Such an $H_{c2}$ agrees reasonably well
with the criterion of zero resistance, achieved at about 6~T, and also
agrees with the $H_{c2}$ values reported previously in thin films of PCCO \cite{Fournier,Alff}
and NCCO \cite{Osofsky,Alff} at $x=0.15$.
A roughly linear growth of $\kappa$ with field is also seen,
for example, in the borocarbide superconductor 
LuNi$_2$B$_2$C \cite{Boaknin2}, 
however with a major difference: in LuNi$_2$B$_2$C, the field induces a 
linear term at $T \to 0$. 
Furthermore, this finite residual linear term ultimately 
reaches a value in the normal state given precisely
by the WF law. In striking contrast, {\sl a magnetic field, even as high as 
13~T, almost twice $H_{c2}$, never induces a residual linear term in PCCO.}  
Given the known value of the electrical
resistivity at $T \to 0$ and 13~T, namely $\rho_0 = 26.8~ \mu \Omega$~cm, the WF law predicts
a residual linear term $\kappa_0/T = L_0/\rho_0 = 0.91$~mW~K$^{-2}$~cm$^{-1}$, as drawn in 
Fig.~\ref{Fig3}.
(Note that weak localization is not expected to alter the WF law appreciably
\cite{Castellani}, 
as verified experimentally in
layered graphene \cite{graphene}.)

This complete violation of the WF law shows that the non-superconducting
ground state of this cuprate material
is not a Fermi liquid. {\sl 
The charge carriers responsible for the good electrical conduction
do not display the expected fermionic heat transport}.

It is instructive to consider the effect of 
increasing the temperature above zero. Contributions to the heat transport from the electron system
($\kappa_e$) and from the ionic lattice ($\kappa_{ph}$) will add to give the measured thermal
conductivity: $\kappa(T) = \kappa_e(T) + \kappa_{ph}(T)$.
These two systems are coupled. It is unambiguous that phonons are scattered by electrons
since the crystal conducts heat better at zero field than at 13 T
beyond 0.6~K.
In other words, as
the temperature is raised, inelastic scattering processes become increasingly
effective
at limiting the phonon 
heat flow, and we expect the dominant process to be electron-phonon scattering.

Given this observed electron-phonon coupling, the total phonon scattering
in field is the sum of whatever scatters phonons in zero field plus the
added scattering that comes from the field-induced electronic excitations.
The thermal resistivity
of phonons due to electron scattering is given by
$W_e^{ph} = B T^{-2}$ \cite{Berman}, where the constant $B$ depends on the strength of the electron-phonon coupling, the electronic density of states, and various other factors. This means
that the phonon conductivity at low $T$ is given by
$\kappa_{ph} = \kappa(0) [ 1 + \kappa(0) W_e^{ph} ]^{-1}$, where $\kappa(0)$ is the 
measured zero-field conductivity, assumed to be essentially all phononic.
%%%% REVISION: corrected spelling of "essentially" %%%%%%%%%%
We can put a reasonable lower bound on the electron-phonon coupling
parameter
$B$ by requiring that the resulting $\kappa_e/T$ does not 
decrease with $T$. Adjusting $B$ such that $\kappa_e/T$ is constant
above 0.3~K gives the curve shown in Fig.~3 for $H = 13$~T (full circles). 
The same can be done for other fields with the resulting values of $\kappa_e/T$ = 0.6,
1.2 and 1.9 mW~K$^{-2}$~cm$^{-1}$ and corresponding values of $B$ = 0.22, 0.4, and 0.68
K$^3$~m~W$^{-1}$ for $H$ = 2, 4, and 8 (or 13) T, respectively. (In
comparison, electron-phonon coupling in copper gives $B = 4-8$~K$^3$~m~W$^{-1}$
\cite{Berman}.) 
Note the self-consistency of this analysis 
whereby $B$ scales with the value of $\kappa_e/T$ (above 0.3 K);
in other words, by increasing the field -- thereby increasing the 
density of electronic excitations -- one increases {\it both} the electronic heat
conduction
and the electron scattering of phonons.

From 
Fig.~\ref{Fig4}, we see that $\kappa_e(T) / T$  exceeds the value of $L_0/\rho$ 
appropriate for a Fermi liquid.
In other words, 
the low-energy excitations of the electron system 
violate the WF law not only at $T \to 0$, where 
$\kappa_e(T) / T \ll L_0/\rho_0$, but also at finite temperatures, where 
$\kappa_e(T) / T > L_0/\rho$ (above 0.2~K). 
It is important to appreciate that qualitatively
the latter violation is independent of our choice of $B$, in the sense that
even if $B$ is set to zero, $\kappa_e/T$ still exceeds $L_0/\rho$, in this case by
30\% at 0.25~K.  We stress that $B=0$ is unphysical, as it would
lead to a 
{\it negative} $\kappa_e/T$ for $T > 0.6$~K.
%%%%%%%% REVISION %%%%%%%%%%
It is also independent of our assumption that $\kappa(0)$ is entirely phononic;
indeed if some part of $\kappa(0)$ is in fact electronic, then the violation turns out to be
even more pronounced. In other words, while there is no accurate way to extract the electronic
contribution at 13 T, the fact that the WF law is violated is independent of our model for the phonon
conduction and the electron-phonon scattering. The result $\kappa_e(T) / T \simeq
2 L_0/\rho_0$ above 0.3~K is obtained in several crystals, even though $\rho(T_c)$  varies in
the range
20 to 40~$\mu \Omega$~cm.

We have investigated the possibility that our measurement gradually ceases to detect the
electronic heat current as electrons fall out of thermal equilibrium with the phonon
bath when the temperature approaches $T=0$. One can imagine this happening 
when the electrical contact resistance between the heater and the sample
is too large and the heat is predominantly carried by the phonons.  This
might then explain the drop in $\kappa_e/T$ measured below 0.3 K or so (see Fig.~\ref{Fig4}).
Two facts lead us to rule out electron-phonon decoupling as the main mechanism for the 
drop. First, identical data is obtained by sending heat directly through the electron system
using photons (as in Ref.~\cite{Decoupling}). Secondly, a similar drop is observed in samples 
of optimally-doped (La,Sr)$_2$CuO$_4$, where we have achieved electrical contact resistances that
are 100 times lower than in our PCCO samples, namely ~ 10~m$\Omega$. 
We therefore believe this low-temperature drop is intrinsic and not
specific to PCCO.\\
%%%%%%%%%%%%%%%%%%%%%%%%
\\
{\bf CONCLUSION}\\

The cuprate PCCO is the first material to violate the Wiedemann-Franz law.
The breakdown of this robust signature of Fermi-liquid theory suggests that the fundamental
entities that carry heat, charge (and spin) in the cuprates are not the usual
Landau
quasiparticles. The fact that the electron system conducts heat in a way which is largely 
unrelated to the way it carries charge at low temperature points to the existence of
neutral excitations responsible for much of $\kappa_e$. If these neutral excitations
are fermions, {\it i.e.} spinons, they should contribute a constant term to $\kappa_e/T$,
as may well be the case for $T > 0.3$~K. 
The drop seen at the very lowest temperatures
then remains to be explained, and the possibility of spinon localization should be considered.

The present study is currently limited to a single carrier concentration, near 
optimal doping, in the whole phase diagram. It will be interesting to investigate 
the evolution with doping of this emerging picture. If one speculates 
that the WF law is satisfied
in overdoped cuprates, where Fermi-liquid theory is generally assumed to hold, our results
on optimally-doped PCCO suggest that a violation of the law is a property of
underdoped cuprates. A comprehensive study will shed light on the gradual breakdown of
Fermi-liquid theory in the cuprates as the carrier concentration is reduced towards
the Mott insulating state.\\
\\
\noindent
{\bf Acknowledgements}\\

We thank C. Lupien, E. Boaknin, M. Sutherland, M. Chiao,
R. Gagnon, J. Brooks, L. Balicas
and B. Brandt 
for their invaluable help in various
aspects of the measurements.
This work was supported by the Canadian Institute for Advanced
Research and funded by NSERC of Canada.  
LT ackowledges the support of a
Premier's Research Excellence Award from the Government of Ontario.
The work in Maryland was supported by the NSF Division of Condensed
Matter Physics.\\
\\
Correspondence should be addressed to L.T. (Louis.Taillefer@utoronto.ca).\\
\\
\noindent
$^\dagger$ Present address:
Laboratoire National des Champs Magn\'etiques Puls\'es,
143 Avenue de Rangueil, BP 4245, 31432 Toulouse, Cedex 04, France. \\
$^\ddagger$ Present address: 
D\'epartement de Physique, Universit\'e de Sherbrooke, Sherbrooke, Qu\'ebec J1K 2R1, Canada.


\begin{references}

\bibitem{Landau} Landau, L.D.  
The theory of a Fermi liquid.
{\it Soviet Physics JETP} {\bf 3}, 920-925 (1957).

\bibitem{BCS} Bardeen, J., Cooper, L.N. \& Schrieffer, J.R. 
Theory of superconductivity.
{\it Phys. Rev.} {\bf 108}, 1175-1204 (1957).

\bibitem{WFL} Wiedemann G. \& Franz R. {\it Annl. Phys.} {\bf 89}, 497-532
(1853).

\bibitem{Sommerfeld} Ashcroft, N.W. \& Mermin, N.D. {\it Solid State Physics} 
(HRW, Philadelphia, 1976), p.~322.

\bibitem{Kearney} Kearney, M.J. \& Butcher, P.N. 
Thermal transport in disordered systems.
{\it J. Phys. C: Solid State Phys.} {\bf 21}, L265-L270 (1988).

\bibitem{Castellani} Castellani, C., DiCastro, C., Kotliar, G., Lee, P.A. \& Strinati, G.
Thermal conductivity in disordered interacting-electron systems.
{\it Phys. Rev. Lett.} {\bf 59}, 477-480 (1987).

\bibitem{Chester} Chester, G.V. \& Tellung, A.
The law of Wiedemann and Franz.
{\it Proc. Phys. Soc.} {\bf 77}, 1005-1013 (1961).

\bibitem{Smrcka} Smrcka, L. \& Streda, P.
Transport coefficients in strong magnetic fields.
{\it J. Phys. C: Solid State Phys.} {\bf 10}, 2153-2161 (1977).


\bibitem{Pobell} 
Gloos, K., Mitschka, C., Pobell, F. \& Smeibidl, P.  
Thermal conductivity of normal and superconducting metals.
{\it Cryogenics} {\bf 30}, 14-18 (1990).

\bibitem{Cu1} Anderson, A.C., Peterson, R.E.  \& Robichaux, J.E.  
Thermal and electrical conductivity of Ag and Pt below 1~K.
{\it Phys. Rev. Lett.}
{\bf 20}, 459-461 (1968).

\bibitem{Cu2} Rumbo, E.R. 
Transport properties of very pure copper and silver below 8.5~K.
{\it J. Phys. F: Metal Phys.} {\bf 6}, 85-98 (1976).

\bibitem{Ag-test} Hill, R.W. \& Boaknin, E. {\it Personal communication}.

\bibitem{Ott} Ott, H.R., Marti, O. \& Hulliger, F.
Low temperature thermal conductivity of CeAl$_3$.
{\it Solid State Commun.} {\bf 49}, 1129-1131 (1984).

\bibitem{Amato} Amato, A. {\it et al}. 
Transport properties of CeCu$_6$ at very low temperature.
{\it J. Magn. Magn. Mater.} {\bf 63 \& 64}, 300-302 (1987).

\bibitem{Suderow} Suderow, H. {\it et al.}
Thermal conductivity and gap structure of the superconducting phases of UPt$_3$.
{\it J. Low Temp. Phys.} {\bf 108}, 11-30 (1997).

\bibitem{Tanatar} Tanatar, M.A. {\it et al.}
Thermal conductivity of superconducting Sr$_2$RuO$_4$ in oriented magnetic fields.
{\it Phys. Rev. B.} {\bf 63}, 064505-1-7 (2001).

\bibitem{Belin} Belin, S. \& Behnia, K.
Thermal conductivity of superconducting (TMTSF)$_2$ClO$_4$: evidence
for a nodeless gap. {\it Phys. Rev. Lett.} {\bf 79}, 2125-2128 (1999).

\bibitem{Pepper} Syme, R.T., Kelly, M.J. \&  Pepper, M. Direct measurement of the thermal conductivity
of a two-dimensional electron gas. {\it J. Phys.: Condens. Matter} {\bf 1}, 3375-3380 (1989).

\bibitem{Kambe} Kambe, S. {\it et al}. Spin-fluctuation mediated thermal conductivity around
the magnetic instability of CeNi$_2$Ge$_2$. {\it J. Low Temp. Phys.} {\bf 117}, 101-112 (1999).

\bibitem{Anderson} Anderson, P.W. 
The resonating valence bond state in La$_2$CuO$_4$ and
superconductivity.
{\it Science} {\bf 235}, 1196-1198 (1987).

\bibitem{Kivelson} Kivelson, S.A., Rokhsar, D.S. \& Sethna, J.P.
Topology of the resonating valence bond state: Solitons and high-T$_c$ superconductivity.
{\it  Phys. Rev. B} {\bf 35}, 8865-8868 (1987).

\bibitem{Fisher} Senthil, T. \& Fisher, M.P.A. 
Z$_2$ gauge theory of electron fractionalization in strongly correlated systems.
{\it  Phys. Rev. B} {\bf 62}, 7850-7881
(2000); {\sl and references therein}.

\bibitem{Millis} Orenstein, J. \& Millis, A.J.  Advances in the physics of high-temperature 
superconductivity. {\it Science} {\bf 288}, 468-474 (2000).

\bibitem{Boebinger} Boebinger, G.S. {\it et al}. 
Insulator-to-metal crossover in the normal state of La$_{2-x}$Sr$_x$CuO$_4$ near optimum doping.
{\it Phys. Rev. Lett.} {\bf 77}, 5417-5420 (1996).

\bibitem{Fournier} Fournier, P. {\it et al}. 
Insulator-metal crossover near optimal doping in Pr$_{2-x}$Ce$_x$CuO$_4$: Anomalous normal-state low 
temperature resistivity.
{\it Phys. Rev. Lett.} {\bf 81}, 4720-4723 (1998).

\bibitem{Growth1} Peng, J.L., Li, Z.Y. \& Greene, R.L. 
Growth and characterization of high-quality single crystals of
R$_{2-x}$Ce$_x$CuO$_{4-y}$ (R=Nd,Sm).
{\it Physica C} {\bf 177}, 79-85 (1991).

\bibitem{Growth2} Brinkmann, M. {\it et al}. 
Crystal growth of high-$T_c$ superconductors 
Pr$_{2-x}$Ce$_x$CuO$_{4+\delta}$ with substitutions of Ni and Co for Cu.
{\it J. Crystal Growth} {\bf 163}, 369-376 (1996).

\bibitem{Mackenzie} Mackenzie, A.P. {\it et al}.   Resistive upper critical field of 
Tl$_2$Ba$_2$CuO$_6$ at low temperatures and high magnetic fields. 
{\it Phys. Rev. Lett.} {\bf 71}, 1238-1241 (1993).

\bibitem{Harus} Harus, G.I {\it et al}. 
Two-dimensional weak localization effects in high-temperature
superconductor Nd$_{2-x}$Ce$_x$CuO$_{4-\delta}$.
{\it Soviet JETP} {\bf 116}, 1-12 (1999).

\bibitem{Fournier2} Fournier, P. {\it et al.} Anomalous saturation of the 
phase coherence length in underdoped Pr$_{2-x}$Ce$_x$CuO$_4$ thin films. 
{\it Physical Review B} {\bf 62}, R11993-R11996 (2000).

\bibitem{Hill} Hill, R.W. {\it et al}. {\it Personal communication}.

\bibitem{Thacher} Thacher, P.D. 
Effect of boundaries and isotopes on the thermal conductivity of LiF.
{\it Phys. Rev.} {\bf 156}, 975-988 (1967).

\bibitem{Taillefer} Taillefer, L. {\it et al}. 
Universal heat conduction in YBa$_2$Cu$_3$O$_{6.9}$.
{\it Phys. Rev. Lett.} {\bf 79}, 483-486 (1997).

\bibitem{Boaknin2} Boaknin, E. {\it et al}. Highly anisotropic gap function in
borocarbide superconductor LuNi$_2$B$_2$C. {\it Phys. Rev. Lett.} accepted
Nov 2001.

\bibitem{Durst} Durst, A.C. \& Lee, P.A.  Impurity-induced quasiparticle transport and universal-limit
Wiedemann-Franz violation in $d$-wave superconductors. {\it Phys. Rev. B} {\bf 62}, 1270-1290 (2000).

\bibitem{Chiao} Chiao, M. {\it et al}. Quasiparticle transport in the vortex state of 
YBa$_2$Cu$_3$O$_{6.9}$. {\it Phys. Rev. Lett.} {\bf 82}, 2943-2946 (1999).

\bibitem{Chiao2} Chiao, M. {\it et al}. 
Low-energy quasiparticles in cuprate superconductors: A quantitative analysis.
{\it Phys. Rev. B} {\bf 62}, 3554-3558
(2000).

\bibitem{Aubin} Behnia, K. {\it et al}. Features of heat conduction
in organic and cuprate superconductors. {\it J. Low Temp. Phys.} {\bf 117}, 1089-1098 (1999).

\bibitem{Taillefer2} Taillefer, L. \& Hill, R.W. Heat transport in high-temperature superconductors.
{\it Physics in Canada}, {\bf 56}, 237-240 (2000).

\bibitem{Lee} Senthil, T. {\it et al}. Quasiparticle transport and localization in high-$T_c$ 
superconductors. {\it Phys. Rev. Lett.} {\bf 81}, 4704-4707 (1998); {\sl and references therein}.

\bibitem{Hussey} Hussey, N.E. {\it et al}.  
Absence of residual quasiparticle conductivity in the underdoped
cuprate YBa$_2$Cu$_4$O$_8$. {\it Phys. Rev. Lett.} {\bf 85}, 4140-4143 (2000).

\bibitem{Sachdev} Vojta, M., Zhang, Y. \& Sachdev, S. 
Quantum phase transitions in d-wave superconductors.
{\it Phys. Rev. Lett.} {\bf 85}, 4940-4943 (2000).

\bibitem{Tsuei} Tsuei, C.C. \& Kirtley, J.R.  
Phase-sensitive evidence for d-wave pairing symmetry in electron-doped cuprate superconductors.
{\it Phys. Rev. Lett.} {\bf 85}, 182-185 (2000).

\bibitem{lambda1} Kokales, J.D. {\it et al}.  
Microwave electrodynamics of electron-doped cuprate superconductors.
{\it Phys. Rev. Lett.} {\bf 85}, 3696-3699 (2000).

\bibitem{lambda2} Prozorov, R.  {\it et al}.  
Evidence for nodal quasiparticles in electron-doped cuprates from penetration depth measurements.
{\it Phys. Rev. Lett.} {\bf 85}, 3700-3702 (2000).

\bibitem{Granath} Granath, M. {\it et al}. 
Nodal quasiparticles and coexisting orders in striped superconductors. 
{\it Phys. Rev. Lett.} {\bf 87}, 167011-1-4 (2001).

\bibitem{Belin2} Belin, S. {\it et al}. 
Probing the upper critical field of $\kappa$-(BEDT-TTF)$_2$Cu(NCS)$_2$.
{\it J. Supercon.} {\bf 12}, 497-500 (1999).

\bibitem{Alff} Kleefisch, S. {\it et al}. 
Possible pseudogap behavior of electron doped high-temperature superconductors.
{\it Phys. Rev. B} {\bf 63}, 100507-100510 (2001).

\bibitem{Osofsky} Osofsky, M.S. {\it et al}.
Temperature dependence of the upper critical magnetic field in BiSrCuO and NdCeCuO.
{\it Proceedings of the 10th anniversary HTS workshop} (World Scientific,
Singapore, 1996), pp.~284-287.

\bibitem{graphene}  Bayot, V. {\it et al}.  
Evidence for weak localization in the thermal conductivity of a 
quasi-two-dimensional electron system. {\it Phys. Rev. Lett.} {\bf 65}, 2579-2582 (1990). 

\bibitem{Berman} Berman, R. {\it Thermal conduction in solids}
(Clarendon, Oxford 1976), p.~145

%%%%%%%% REVISION %%%%%%%%%%

\bibitem{Decoupling} Gutsmiedl, P., Probst. C. \& Andres, K.
Low temperature calorimetry using an optical heating method.
{\it Cryogenics}  {\bf 31}, 54-57 (1991).

%%%%%%%%%%%%%%%%%%%%%%%%%

\end {references}
\end{document}